# RAP-NET: COARSE-TO-FINE MULTI-ORGAN SEGMENTATION WITH SINGLE RANDOM ANATOMICAL PRIOR

*Ho Hin Lee [1], Yucheng Tang [1], Shunxing Bao [1], Richard G. Abramson [2], Yuankai Huo [1], Bennett A. Landman [1,2]*

[1]Electrical Engineering and Computer Science, Vanderbilt University, Nashville, TN USA
[2]Radiology and Radiological Sciences, Vanderbilt University Medical Center, Nashville, TN, USA

## ABSTRACT

Performing coarse-to-fine abdominal multi-organ segmentation facilitates to extract high-resolution segmentation minimizing the lost of spatial contextual information. However, current coarse-to-refine approaches require a significant number of models to perform single organ refine segmentation corresponding to the extracted organ region of interest (ROI). We propose a coarse-to-fine pipeline, which starts from the extraction of the global prior context of multiple organs from 3D volumes using a low-resolution coarse network, followed by a fine phase that uses a single refined model to segment all abdominal organs instead of multiple organ corresponding models. We combine the anatomical prior with corresponding extracted patches to preserve the anatomical locations and boundary information for performing high-resolution segmentation across all organs in a single model. To train and evaluate our method, a clinical research cohort consisting of 100 patient volumes with 13 organs well-annotated is used. We tested our algorithms with 4-fold cross-validation and computed the Dice score for evaluating the segmentation performance of the 13 organs. Our proposed method using single auto-context outperforms the state-of-the-art on 13 models with an average Dice score 84.58% versus 81.69% (p<0.0001).

## 1. INTRODUCTION

Creating a robust and accurate pipeline for volumetric abdominal organ segmentation is challenging. Deep neural networks have been used for performing sematic segmentation in medical imaging perspective with 2D images or 3D patches, such as DeepLab [3], UNet [4] and VoxResNet [5]. Current major challenges of abdominal organ segmentation present in three characteristics: 1) weak intensity boundaries between abdominal organs, 2) large morphological variation of different organs, and 3) high resolution of 3D volumes.

To overcome the limitation of using 2D CNNs, 3D volumes are sliced in axial, coronal and sagittal directions and used for coarse organ detection and perform segmentation with the detection output [6]. Both 2D and 3D-patch based learning methods target the single organ approaches, and it is still challenging to extend to the whole abdominal interest regions. *Roth et al.* proposed coarse-to-fine pipeline with scaling input volumes at different levels using multi-scale pyramid networks and compute refine prediction at the last selective level [1]. *Zhu et. al.* added expanded bounding box into coarse-to-fine pipeline to abstract the ROI of small targets and improve robustness for refining segmentation [2]. However, numerous organ-corresponding models are needed to be trained to obtain refine segmentation performance. Significant effort is allocated for tuning hyperparameters, and there is a lack of flexibility to combine contextual information across scales due to high complexity of training strategies. Backpropagating multiple loss functions for all organs to obtain a single model is not possible due to the missing organ regions in the extracted patches. Therefore, a single high-resolution refine model framework integrating global and local contextual information to refine segmentation is needed.

To resolve above critical challenges, we introduce a 3D hierarchical coarse-to-fine framework with a single model in the refine level for multi-organ segmentation. Briefly, we initially down-sampled the volumetric images and used a traditional approach to generate a coarse segmentation for each organ. The coarse segmentation then acts as an anatomical prior for each organ and extracted corresponding organ patches using the prior information. We integrate the corresponding organ prior to patches as the second channel input with the image patches and train the refine model with all binary labeled organ patches. The refined model can then encode the variability of shape and local intensity across all organs and limit the segmentation region with the anatomical prior, generating a robust and accurate performance for multi-organ segmentation.

## 2. MATERIALS AND METHODS

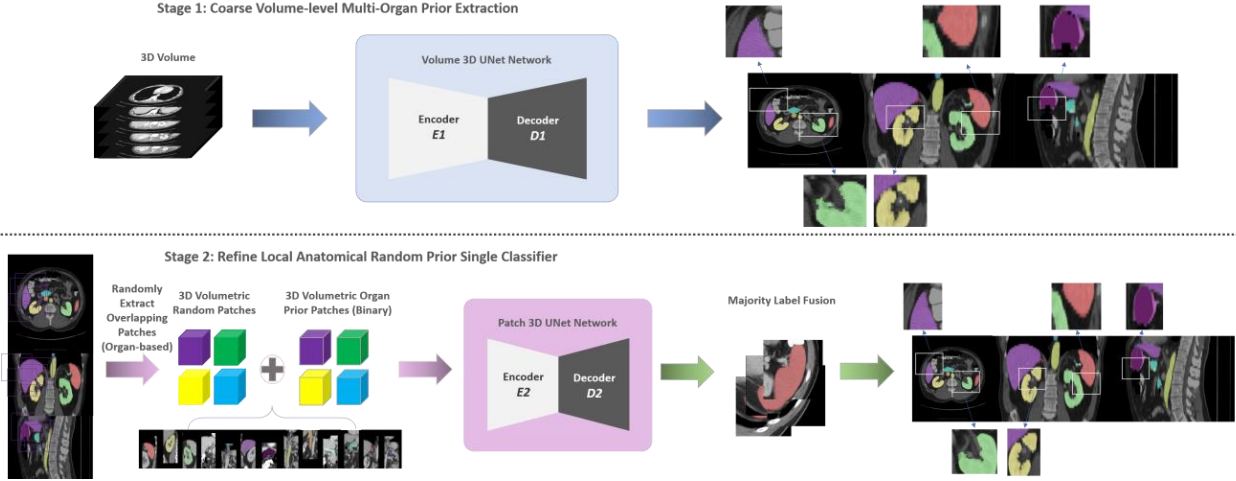

**Fig. 1.** Overview of the pipeline combining the global and local level representations. The proposed hierarchical coarse-to-fine multi-organ segmentation framework consists of two main stages: 1) coarse global anatomical multi-organ prior extraction and 2) a refined single classifier with local anatomical random prior. The coarse- and refined- model are end-to-end optimized separately. The predicted segmentation patches of each single organ are finally fused as multi-channel volume and converted to multi-organ segmentation mask with majority voting.

Our work aims to create a single model adapting high-resolution multi-organ context across global and local levels. As shown in Fig. 1, the backbone of the hierarchical approach is based on the state-of-the-art [7], which consists of 13 refined organ models separately capturing the local representation of each corresponding labeled organs. We propose a refined organ-voxel classifier that maintain the ability of classifying the morphological variations and abstracting the local contrast characteristics across all organs. By utilizing the anatomical prior information extracted from all organs, the extracted local representation from the single patch-wise multi-organ segmentation model integrates all organs' local intensity feature and the global morphological information, to generalize the structural variability and eliminate the possibility of over-segmenting towards neighboring organs. The proposed code is available at https://github.com/MASILab/coarse_to_fine_prior_seg.

**2.1. Data**

A clinical research cohort with 100 patient volumes in portal venous phase was retrieved in de-identified form under the approval of the local IRB (institutional review board). The range of slice numbers across all volumes in the cohort is between 42 and 149 with a dimension of $512 \times 512$. The resolution for x, y and z-axis are in the range of 0.5-0.9 mm, 0.5-0.9 mm and 2.5-7.0 mm, respectively. We first randomly split the dataset into 80 volumes as the training and validation set, and the other 20 volumes as the testing dataset. Each volumetric scan is manually annotated with 13 classes of multiple abdominal organs including spleen, right and left kidneys, gall bladder, esophagus, liver, stomach, aorta, inferior vena cava (IVC), portal splenic vein (PSV), pancreas, right and left adrenal glands.

We initially down-sampled each volume to a resolution of $2 \times 2 \times 6$ mm and pad/crop to a constant dimension of $168 \times 168 \times 64$ as the input for the coarse stage segmentation model. The predicted multi-organ segmentation mask of each volume is then converted back to its original corresponding resolution and extracts volumetric patches with the coarse segmented mask in dimension of $128 \times 128 \times 64$ as the input for the refined stage segmentation model. The single refine model generates a 3D binary mask, corresponding to the organs extracted from the anatomical prior.

**2.2. Global anatomical multi-organ prior extraction**

We first adapt a volume-based 3D UNet architecture as the base of our network and provide supervisory end-to-end optimization to achieve a coarse level multi-organ segmentation [8]. The modified network consists of 8 encoders with convolutional kernel size of $3 \times 3 \times 3$, batch normalization layers, and 10 decoders with deconvolutional kernel size of $2 \times 2 \times 2$. Skip connections are used to integrate and capture the small variant representations from encoder blocks. ReLU activation units are used in both encoder and decoder blocks. The global representations can then be abstracted among all organs with the integration of high-level features from encoder to decoder. For multiple classes of abdominal organs $A$, we define the output of the final layer from 3D UNet as $d_0 \in R^{(C-1) \times S \times H \times W}$, where H, W, S and C are the number of height, width, slices and the channel number of the predicted multi-organ segmentation. A softmax activation is used to compute the probability map of the predicted segmentation $p(A) = softmax(d_0)$ for each voxel. Each value of $p(A)$ is extracted and compared with the similarity with the ground truth multi-organ label with multi-source Dice loss (MSDL).

$$MSDL = -\frac{2}{A} \frac{\sum_{a=0}^{A} w \sum_{i=1}^{M} \sum_{i=1}^{N} V_{ij} S_{ij} + \emptyset}{\sum_{a=0}^{A} w \sum_{i=1}^{M} \sum_{i=1}^{N} V_{ij} + \sum_{a=0}^{A} w \sum_{i=1}^{M} \sum_{i=1}^{N} S_{ij} + \emptyset} \quad (1)$$

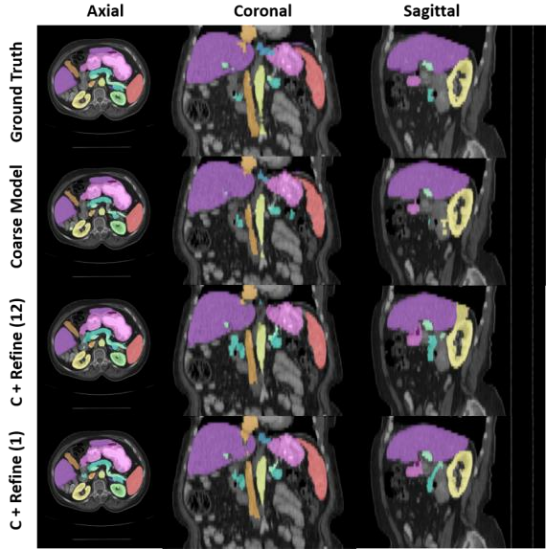

**Figure 2.** The qualitative representation of the multi-organ segmentation result with coarse level model, coarse (C) and 13 refine organ-specific (R) models, and coarse and 1 single refine model.

where *A* is the number of organ anatomies, *w* represents the variance between labels set properties. *S* and V provides the segmentation probability and the intensity of voxel belongs to the classes of organs respectively. A function ∅ is created to compute the correlation of the prediction and voxel value.

### 2.3. Single classifier with local anatomical random priors

To capture the local variation in feature representation, 50 volumetric patches are extracted randomly with the corresponding anatomical prior extracted from the coarse model to ensure the overlapping region covering the complete volume of organ. In total, 39000 organ patches are selected and directly used for end-to-end single model training. The extracted patches include a certain amount of neighboring organ voxels apart from the main organ and lead to the adverse effect of segmenting main organ in corresponding patches. Here, we integrate the global anatomical prior with the volumetric patches as two channels input for training and adapt the 3D UNet model architecture with the same network configuration of the coarse segmentation model [8]. However, we modified the number of classes label in patches. Only the corresponding organ patch label is being extracted as binary mask using anatomical organ priors as the label input to the refine model. We define the output of the final layer from 3D UNet model as $d_1 \in R^{(2) \times S \times H \times W}$, where H, W, S and 2 are the number of height, width, slices and the number of channels referring to the background and specific organ.

Training as single model framework provides an opportunity to adapt significant variation of the morphological and contrastive characteristics from large organs to small organs. Large numbers of organ-corresponding patches for training increases the generalizability of the model to segment all organs with prior information. As binary labels are used for all organs, the shape variation from small to large organs are also adapted to the feature representation for model to encode. With the integration of morphological and intensity variation characteristics across organs, the abstracted representations increase the localization ability of model. The integration of anatomical prior preserves the global anatomical location and the boundaries of specific organs in abdominal regions, while the random prior patches capture the large variability of shape and voxel intensities in the local regions, generating connective linkage between the voxel intensity variability and the morphological characteristics to stabilize the segmentation performance with the integrated representations from all organs.

Single channel Dice loss is used as the loss function for optimizing the binary segmentation for organ patches. The predicted segmentation for each patch is a binary mask corresponding to the specific organ. All binary masks from each organ patch are fused to generate a 13 channel multi-organ segmentation mask with majority voting.

### 2.4. Implementation details

We performed 4-fold cross-validation with our labeled clinical cohorts to ensure both coarse and refine level model can capture the anatomical information of each organs. In total, 80 volumes of the clinical research cohorts are used for training and validation. 80 volumes are randomly shuffled and split to 4 groups of combinations with 60 volumes for training and 20 volumes for validation. The optimized model is chosen with the best validation performance for segmentation across all folds. The testing cohort is the BTCV MICCAI 2015 Challenge testing dataset with 20 volumes. The batch size was set to 1 for coarse volume-based model, while it was set to 2 for refine patch-based model. Adam was used as the optimizer for both stages end-to-end training. We first trained the coarse segmentation model with 100 epochs with learning rate of 0.0001 and choose the model with the lowest validation loss. For the single refine model, we directly input 39000 patches and trained the model for 5 epochs with learning rate of 0.0001. To evaluate the segmentation performance of both models, Dice score is used as the evaluation metric and compute a quantitative measure of the overlapping similarity between the prediction and the ground truth label. Subject volumes without gall bladder are eliminated in calculating the quantitative measures of gall bladder organ only.

### 3. RESULTS AND DISCUSSION

Table 1 shows the quantitative comparison of the segmentation performance with *Roth et. al.*, the state-of-the-art method, *Zhu et. al.* and our proposed model. The average Dice coefficient of all organ segmentation is increased from 81.69% to 84.58% and the standard deviation of Dice decreased from an average of 13.0% to 11.5% comparing to the state-of-the-art. We also compared our pipeline with the *Zhu*'s and *Roth*'s method. Our proposed pipeline demonstrated a better segmentation performance than *Zhu*'s method in most of the organs. The anatomical prior information provided a localization impact for improving the segmentation result. The single model encodes the variation across all organs with all organ patches directly training, while the 13 models only encode with the corresponding organ patches, the combination of intensity and morphology feature boost up the performance in the single model.

| Methods | Spleen | R Kid. | L Kid. | Gall. | Eso. | Liver | Stomach | Aorta | IVC | PSV | Pancreas | RAD | LAD | AVG |
|---|---|---|---|---|---|---|---|---|---|---|---|---|---|---|
| C only | 0.921 | 0.828 | 0.894 | 0.695 | 0.670 | 0.935 | 0.783 | 0.885 | 0.814 | 0.627 | 0.690 | 0.604 | 0.615 | 0.770 |
| Roth [1] | 0.926 | 0.884 | 0.889 | 0.531 | 0.724 | 0.953 | 0.819 | 0.884 | 0.823 | 0.687 | 0.720 | 0.664 | 0.693 | 0.784 |
| C+R (13) | 0.939 | 0.900 | 0.943 | 0.763 | 0.712 | 0.952 | 0.822 | 0.897 | 0.828 | 0.710 | 0.745 | 0.646 | **0.754** | 0.817 |
| Zhu [2] | 0.961 | **0.928** | 0.932 | 0.693 | 0.772 | **0.964** | **0.849** | 0.913 | 0.837 | 0.698 | 0.762 | 0.684 | 0.721 | 0.824 |
| C+R (1) | **0.965** | 0.920 | **0.945** | **0.793** | **0.783** | 0.960 | 0.833 | **0.916** | **0.856** | **0.762** | **0.766** | **0.741** | 0.746 | **0.846** |

**Table 1.** Dice score comparison of the testing cohort between coarse model, *Roth et. al.*, coarse + 13 refine models (state-of-the-art), *Zhu et. al.* and our proposed model. Our proposed pipeline outperformed the other three coarse-to-fine methods with an average Dice of 84.58% (p<0.0001, Wilcoxon signed-rank test). C is short for coarse and R is short for refine. The number next to R refer to the number of refine models.

Fig. 2 further demonstrated the confidence level of the quantitative result with qualitative representations. The segmentation result computed from coarse model can provide approximate anatomy localization information and the refine model can well adapt the organ regions for refining segmentation. The 13 organ corresponding models is well adapted with the anatomical information of the corresponding organ from the prior extracted in the coarse stage and abstracted the local intensity feature of each organ to refine the single target organ segmentation with the corresponding organ binary label. However, they only capture the corresponding morphological variation of the single target organ and learn the specific contrast characteristics within the prior preserved region, computing segmentations separately without anatomical linkage between organs. Over-segmenting the neighboring organs still existed due to the similarity of neighboring organ voxel intensity. After we trained as a single model, the neighboring organ boundaries are substantially identified, as it encoded all organs' variation of the morphological changes in anatomical prior. The single refine framework provides an opportunity to control the refine segmentation in specific organ locations integrating with other organ localization information.

## 4. CONCLUSIONS

The proposed coarse-to-fine framework allows a single deep learning model to encode the integration of morphological and contrastive characteristics with multiple abdominal organs. Further studies will be performed by inputting multi-organ priors instead of one single corresponding organ prior, as other channels. The morphological linkage between organs can be evaluated in the future and innovate stratifying approach according to the anatomical characteristics between abdominal organs.

## 5. ACKNOWLEDGMENTS

This research is supported by HuBMAP, ULTR00045, NSF CAREER 1452485 and NIH 1R01EB017230 (Landman). We gratefully acknowledge the support of NVIDIA Corporation with the donation of the Titan X Pascal GPU usage and Dr. Michael R. Savona. There are no conflicts of interest.

## 7. COMPLIANCE WITH ETHICAL STANDARDS

This study was performed in line with the principles of the Declaration of Helsinki. Approval was granted by the IRB (institutional review board) of Vanderbilt University.